\RequirePackage{lineno} 

\documentclass[12pt]{iopart}
\usepackage{graphicx}
\def\de{\delta}

\def\th{\theta}

\def\De{\Delta}

\def\nue{\nu_e}
\def\numu{\nu_\mu}

\def\nuebar{\bar\nu_e}
\def\numubar{\bar\nu_\mu}

\newcommand{\beq}{\begin{eqnarray}}
\newcommand{\eeq}{\end{eqnarray}}
\def\to{\rightarrow}

\def\no{\nonumber}

\def\ueVt{\mbox{eV}^2}

\begin{document}
\nocite{*}

\title{Short Baseline Neutrino Oscillation Experiments}


\author{Teppei Katori}

\address{Queen Mary University of London, London E1~4NS, UK}

\ead{t.katori@qmul.ac.uk}

\begin{abstract}
Series of short baseline neutrino oscillation experiments provided unexpected results, 
and now they are called “{\it short baseline anomalies}”, and all indicates 
an existence of sterile neutrinos with a mass scale around 1~eV.  
The signals of short baseline anomalies are reported from 4 different classes of experiments. 
However, at this moment, there is no convincing theoretical model 
to explain such sterile neutrinos, 
and a single experiment to confirm 1~eV sterile neutrinos may be challenging. 
In this short note, we describe classes of 
short baseline neutrino oscillation experiments and their goals. 
 
\end{abstract}

\section*{Classification of short baseline neutrino oscillation experiments}

The short baseline anomalies come from 4 different classes of experiments~\cite{sterile}. 
Based on this, we can classify short baseline neutrino oscillation experiments into 
following 5 groups;
\begin{enumerate}
\item test of LSND signal, 
\item test of MiniBooNE signal, 
\item test of reactor antineutrino anomaly,
\item test of Gallium anomaly, and
\item others.
\end{enumerate}
We discuss each of the above group in following sections. 
The last group is all other experiments, 
they are mainly experiments motivated by 1~eV sterile neutrinos 
and not short baseline anomalies themselves. 
Therefore many experiments in (v) are not short baseline oscillation experiments. 

The short baseline anomalies are unsolved mysteries in this community, 
and they attract many theorists and experimentalists. 
The search of 1~eV sterile neutrino is one of the big branches of 
the neutrino experiment community~\cite{neutrino}. 
Therefore, it is rather impossible to cover all experiments in this note, 
however, we try to cover most of experiments planned in the near future. 

\section{LSND signal and experiments designed to test to it}

\subsection*{LSND experiment}

The origin of $\De m^2_{sterile}\sim 1~\ueVt$ is the LSND experiment~\cite{LSND_osc}, 
where muon antineutrino (0 to 53~MeV) are produced by pion decay-at-rest (DAR), 
and detected by a liquid scintillator detector at 31~m from the target. 
The LSND experiment measured $\nuebar$ candidate events by 
utilizing the coincidence of the prompt Cherenkov radiation 
from the positron and the delayed neutron capture by a hydrogen.
\beq
\numubar\stackrel{oscillation}{\longrightarrow}\nuebar+p\to\e^+(Cherenkov)+n(capture)~.\no
\eeq 

The LSND experiment observed an excess of $\nuebar$ candidate events.  
This small ($<1\%$ oscillation probability) 
but statistically significant signal is 
consistent with the presence of sterile neutrinos ($\numubar\to\nu_{sterile}\to\nuebar$). 

Meantime, the KARMEN experiment~\cite{KARMEN} excluded high $\De m^2$ region, 
and the Bugey experiment~\cite{Bugey} excluded all low $\De m^2$ region 
of the LSND signal region in $\De m^2 - sin^2 2\th$ plane. 
The combined result suggests the LSND signal is most likely due to sterile neutrinos around 
1~eV region.

\subsection*{Experiments to test LSND signal}

The LSND experiment has  limited statistics, 
also, the duty cycle (nominal run, $\sim$25~Hz) was high with a wide pulse. 
This allows LSND to accept large amount of cosmic backgrounds. 
Also, the detector was located to the direction of the primary beamline, 
and neutrinos from pion decay-in-flight (DIF) made additional backgrounds. 
Therefore, to test LSND signal, experiments are desired to have;
\begin{itemize}
\item LSND beam energy, baseline, and the detector which can tag $\nuebar$ candidate events, 
\item higher statistics, 
\item known and narrow beam structure, and 
\item detector located on off-axis. 
\end{itemize}

The promising experiments are off-axis liquid scintillator experiments 
at various neutron spallation sources in the world. 
The high pion-DAR $\numubar$ flux is available (and free!) 
from Oak Ridge National Laboratory (ORNL)~\cite{OscSNS},  
J-PARC Materials and Life Science Experimental Facility (MLF)~\cite{OscJSNS}, 
and European spallation source (ESS)~\cite{ESSnuSB}. 
On top of the high neutrino flux, proton pulses hitting the target 
to produce neutrons have well known beam structure. 
Therefore these experiments cover the desired features to test the LSND signal. 
Presently, OscSNS~\cite{OscSNS} is about to write a proposal, 
and J-PARC group~\cite{OscJSNS} submitted proposal to J-PARC.

\section{MiniBooNE signal and experiments designed to test it}

\subsection*{MiniBooNE experiment}

The MiniBooNE experiment is designed to test the LSND signal within  
the two massive neutrino oscillation hypothesis. 
However, muon (anti)neutrinos are now made by 
pion DIF at the Fermilab Booster neutrino beamline~\cite{MB_beam}, 
and the baseline is 541~m from the target (pion decay length is $\sim$18~m). 
The MiniBooNE detector is a spherical mineral oil based Cherenkov detector~\cite{MB_detec}, 
and the $\nue$($\nuebar$) candidate signals are 
measured as single isolated electron-like Cherenkov ring.
In this way, the systematics of MiniBooNE 
is completely different from the LSND experiment, 
but MiniBooNE can test 1~eV sterile neutrino hypothesis, 
because of similar L/E with LSND. 
\beq
\numu&\stackrel{oscillation}{\longrightarrow}&\nue+n\to p+e^+(Cherenkov)~,\no\\
\numubar&\stackrel{oscillation}{\longrightarrow}&\nuebar+p\to n+\e^-(Cherenkov)~.\no
\eeq 

The Booster neutrino beamline can run either in neutrino mode or in antineutrino mode, 
by focussing either positive or negative mesons. 
Since LSND signal was interpreted $\numubar\to\nu_{sterile}\to\nuebar$ oscillations, 
running in antineutrino mode is more interesting. 
However, the antineutrino mode run suffers from lower statistics 
and higher backgrounds~\cite{MB_ANTICCQE} 
(especially from the muon-neutrino contamination in the antineutrino mode beam~\cite{MB_WS}), 
therefore the experiment started in neutrino mode prior 
to the antineutrino mode running, 
and in the meantime systematics (the neutrino flux, neutrino interactions, 
and the detector response) were studied. 

Unlike the LSND experiment, expected signal to noise is much lower. 
There are 2 dominant backgrounds of $\nue$($\nuebar$) candidate events. 
The first one is the intrinsic $\nue$($\nuebar$) contaminated in the beam. 
The majority of them are made by muon decay, therefore, 
MiniBooNE constrains them by simultaneously measuring 
$\numu$ charged current quasi-elastic (CCQE) events~\cite{MB_CCQE,MB_CCQEPRD}, 
where measured $\numu$ is related to intrinsic $\nue$ through  
the pion decay chain ($\pi^+\to\numu+\mu^+,~\mu^+->\numubar+e^++\nuebar$) in their simulation. 

The second largest background is the misID of neutral current (NC) events, 
mainly NC $\pi^{\circ}$ production. 
Although a $\pi^{\circ}$ decays to two gamma rays which should be distinguishable from 
an electron (positron) Cherenkov ring, 
sometimes decay kinematics make two gamma rays look like one gamma ray 
(asymmetric decay, gamma rays are too close). 
Then, the Cherenkov ring from one gamma ray is indistinguishable from an electron (positron). 
For this, MiniBooNE internally measured NC$\pi^{\circ}$ production, 
and the measured information was used to correct $\pi^{\circ}$ production rates 
in the simulation~\cite{MB_pi0}.

\begin{figure}[t]
\begin{center}
\includegraphics[scale=0.5]{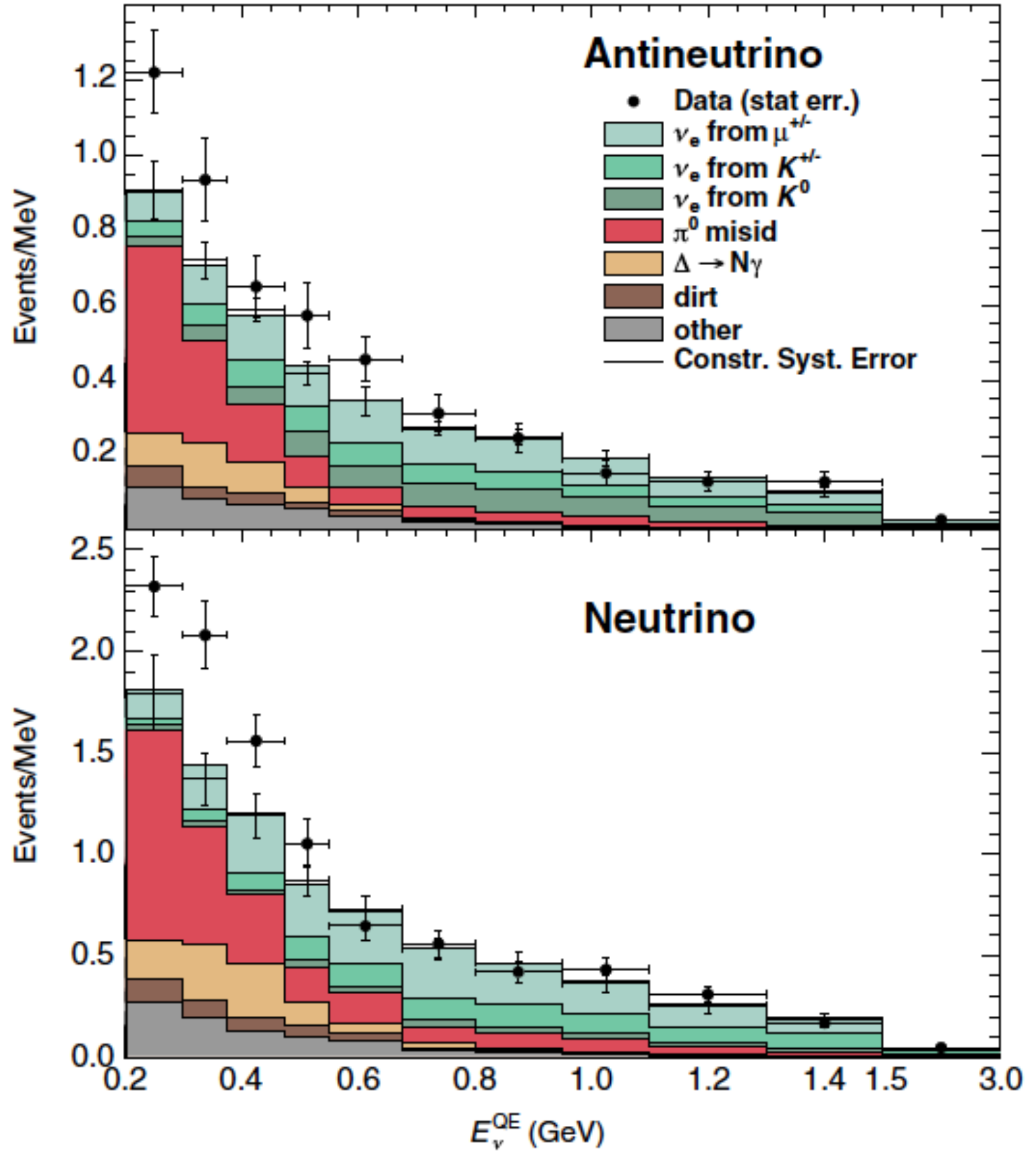}
\includegraphics[scale=0.5]{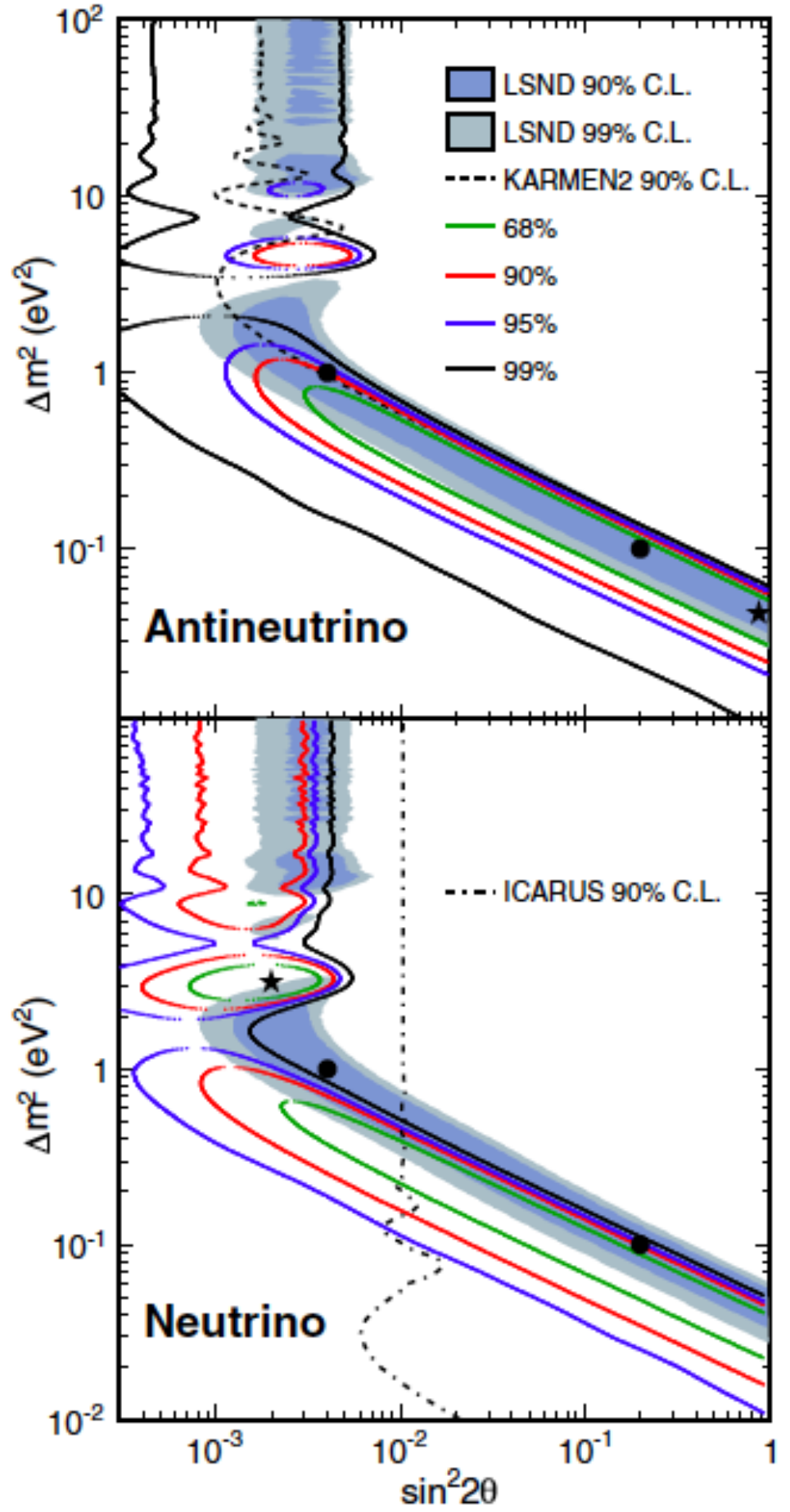}
\end{center}
\caption{\label{fig:MB_osc}
(color online) 
The final MiniBooNE oscillation results~\cite{MB_osc}. 
The top plots are antineutrino mode results, 
and bottom plots are neutrino mode results.
The left plots show the reconstructed neutrino energy distribution 
of oscillation candidate events, 
and the right plots show the allowed region in $\De m^2$-$sin^22\th$, 
where the best fit points are shown in black stars. 
Both modes show excesses in the low energy region, 
while the neutrino mode has higher statistical significance. 
On the other hand, 
the compatibility with the LSND signal is better in antineutrino mode.
 }
\end{figure}

After the 10 years running in both neutrino and antineutrino mode,  
the MiniBooNE experiment observed excesses in both neutrino 
and antineutrino mode runs~\cite{MB_osc}. 
The final result is shown in Figure~\ref{fig:MB_osc}.

\subsection*{Experiments to test MiniBooNE signal}

The measured signal, especially in neutrino mode, 
does not quite agree with the expected sterile neutrino signal. 
The MiniBooNE detector cannot distinguish an electron (positron) and a gamma ray, 
therefore $\numu$NC interaction 
with single gamma ray in the final state is a potential misID background. 
Therefore, to test the MiniBooNE signal, experiments are desired to have;
\begin{itemize}
\item MiniBooNE beam energy and baseline,  
\item ability to distinguish NC or CC interaction, or 
\item ability to distinguish an electron (positron) and a gamma ray.
\end{itemize}

The MiniBooNE+ was proposed to fulfil these criteria~\cite{MB+}. 
By doping scintillator (PPO) in the MiniBooNE detector mineral oil, 
MiniBooNE+ can measure scintillation light from the neutron capture. 
This allows statistical separation between $\nue$CCQE interaction 
(higher proton multiplicity in the final state), 
and $\numu$NC interactions 
(protons and neutrons are half-and-half in the final state). 
However, the proposal of MiniBooNE+ was not accepted by Fermilab recently.

The MicroBooNE experiment~\cite{uB} is a new experiment on 
the Fermilab Booster neutrino beamline to test the MiniBooNE signal. 
It is also an important project for the future large liquid argon (LAr) TPC experiment,
such as LBNE~\cite{LBNE}. 
The MicroBooNE LArTPC detector 
has an ability to separate an electron from single gamma ray, 
by utilizing vertex-shower separation and dE/dx before developing the shower. 
This clearly tells if MiniBooNE excess is by an electron (positron) or a 
gamma ray~\cite{Fleming_uB}. 
The MicroBooNE experiment is under commissioning stage, 
and they expect beam data at the end of 2014.
  
Although the T2K experiment~\cite{T2K_osc} is designed to measure 
the neutrino Standard Model ($\nu$SM) parameters and it does not use the Booster neutrino beam, 
J-PARC neutrino beam~\cite{T2K_beam} has a similar beam peak ($\sim$600~MeV) 
as the Booster neutrino beam but is narrower, 
and the baseline to the near detector complex~\cite{T2K_detec} 
is close (280~m) to what MiniBooNE has.  
Therefore, T2K is sensitive to the MiniBooNE signals~\footnote{
Current sterile neutrino search analysis in T2K is looking for 
$\nue$ disappearance~\cite{T2K_sterile}, instead of $\nue$ appearance.}.   
The magnetic field in the near detector is a great advantage.  
It can allow the sign selection of the signal. 
The NC background (mostly ambient gamma rays) can be understood 
from the internal measurement~\cite{T2K_nue}. 

\section{Reactor antineutrino anomaly and experiments designed to test it}

\subsection*{Reactor antineutrino anomaly}

The re-evaluation of reactor electron antineutrino flux calculation provides 
consistently lower rate than world reactor data (about 6\%)~\cite{ReactorAnomaly1}. 
This, so called {\it reactor antineutrino anomaly} 
can be interpreted as neutrino oscillations with 
1~eV sterile neutrinos~\cite{ReactorAnomaly2}. 

\subsection*{Experiments to test Reactor anomaly}

Reactor anomaly can be tested by small scale experiments,  
by measuring neutrino flux with small detectors with very short baseline ($\sim$15~m). 
To detect low energy reactor antineutrino ($\sim$4~MeV), 
detector needs to be sensitive to low energy events. 
The common choice is the liquid scintillator detector, 
where large mass can be prepared at a relatively low cost. 
 
There are number of such experiments designed for R\&D of 
neutrino reactor monitoring for 
nuclear non-proliferation (SCRAAM~\cite{SCRAAM}, Nucifer~\cite{Nucifer}, etc).
These experiments are naturally served to test reactor antineutrino anomaly. 
Due to affordable cost of the experiments, 
several new experiments are also proposed to test the reactor antineutrino anomaly 
(DANSS~\cite{DANSS}, PROSPECT~\cite{PROSPECT}, STEREO~\cite{STEREO}, etc).

\section{Gallium anomaly and experiments designed to test}

\subsection*{Gallium anomaly}

The two of pp-solar neutrino experiments, SAGE~\cite{SAGE} and GALLEX~\cite{GALLEX}, 
used highly radioactive sources to calibrate gallium detectors.
\beq
\nue + ^{71}Ga \to ^{71}Ge + e^-~. 
\eeq 
But some of these measurements using mega-curie $^{51}$Cr or $^{37}$Ar sources 
showed lower event rates than expected, 
and this so-called {\it Gallium anomaly} 
can be understood by neutrino oscillations with 1~eV sterile neutrinos~\cite{Giunti_Gallium}.

\subsection*{Experiments to test Gallium anomaly}
 
To test “Gallium anomaly”, a highly radio-active neutrino source 
and a very sensitive detector are required. 
Existing high sensitivity solar or reactor neutrino detectors 
(Borexino as ``SOX''~\cite{SOX}, KamLAND as ``Ce-LAND''~\cite{Ce-LAND}, 
SNO+~\cite{SNO+}, DayaBay~\cite{DayaBay+}, etc) 
are good candidates for this purpose.
Similarly, SAGE group proposed to build a new detector but 
reuse the liquid gallium from old detectors~\cite{SAGE+}.   
Proposed solar neutrino experiment (LENS~\cite{Lens-sterile}, etc) and coherent scattering 
experiment (RICOCHET~\cite{RICOCHET}, etc) can look for 1~eV sterile neutrinos, too.  

\section{Others, 1~eV sterile neutrino searches}

\subsection*{Tests by existing facilities}

Once we assume 1~eV sterile neutrinos, some existing facilities are also sensitive to signals, 
even though the experiments are not originally designed to test 1~eV sterile neutrinos. 
The IceCube experiment~\cite{IceCube_PeV} is designed 
to measure astrophysical ultra high energy neutrinos, 
however, 1~eV sterile neutrinos cause disappearance signals  
for $>$100~GeV high energy atmospheric neutrinos~\cite{IceCube_sterile}. 
MINOS+ experiment~\cite{MINOS+} is an extension run of 
MINOS experiment~\cite{MINOS_osc,MINOS_sterile} during NOvA beam configuration era 
(medium energy NuMI, $\sim$7~GeV peak at Sudan mine)~\cite{NOvA}. 
One of physics goal of MINOS+ is to look for 
$\numu$ disappearance signal due to 1 eV sterile neutrinos. 

\subsection*{Tests by R\&D facilities}

Many R\&D experiments for other purposes often look for 1~eV sterile neutrinos. 
For example IsoDAR experiment~\cite{IsoDAR} look for sterile neutrino oscillation using 
$^8$Li isotope made by the high power cyclotron.  
This cyclotron is a part of the R\&D for the DAE$\de$ALUS experiment~\cite{DAEdALUS}. 
The $\nu$STORM~\cite{nuSTORM} experiment uses the muon storage ring, 
which is an important step for the future neutrino factory~\cite{VLENF}. 
KDAR experiment~\cite{KDAR} uses mono-energetic kaon DAR muon neutrinos (236~MeV), 
and the detector requires high resolution, such as LArTPC. 
This can be considered a part of LArTPC technology R\&D,  
and in fact, all LArTPC sterile neutrino searches, such as MicroBooNE~\cite{uB}, 
LAr1-ND~\cite{LAr1-ND}, and NESSiE~\cite{NESSiE}, 
have detector R\&D aspects for future large LArTPC experiments. 

\subsection*{Ultimate 1~eV sterile neutrino search experiments}

Experiments including precise measurement of oscillation probability with 
function of L/E can be considered in this group. 
LSND reloaded~\cite{LSND-reloaded} was proposed to test short baseline neutrino oscillations by 
measuring oscillations with function of L/E in a large detector, 
such as gadolinium doped Super-Kamiokande. 
Similar concept may be applied to reactor antineutrino anomaly experiments 
and gallium anomaly experiments, 
where neutrino sources are small and low energy. 
In those experiments, precise L/E dependence measurement may be possible by 
either using a large detector or multiple small detectors, 
or moving sources and/or detectors.
The precise oscillation measured in this way 
is a strong evidence of sterile neutrinos 
and it is a missing part from past experiments. 

\section*{Acknowledgement}

The author thanks Ranjan Dharmapalan for the careful reading of this manuscript.  
The author thanks to the organisers of ``NuPhys2013, prospects in neutrino physics, 
(Institute of Physics, London, UK)'' for the invitation to the conference.

\section*{References}
\bibliography{TK_SBL_NuPhys2013}

\end{document}